\documentclass[aps,prd,twocolumn,showpacs,superscriptaddress,amsmath,graphicx,psfrag,color,longtable]{revtex4}

\usepackage{psfrag}
\usepackage{epsfig}
\usepackage{color}
\usepackage{longtable}
\usepackage{braket}

\definecolor{Red}{rgb}{1.,0.,0.}
\newcommand{\Red}[1]{{\color{Red}{#1}}}

\begin{document}

\title{Chiral corrections to the scalar form factor in $B_{q} \to D_{q}$ transitions}

\author{Jan O. Eeg}
\affiliation{University of Oslo, Physics Department, P. O. Box 1048 Blindern, N-0316 Oslo 3, Norway}

\author{Svjetlana Fajfer}
\email[Electronic address:]{svjetlana.fajfer@ijs.si}
\affiliation{J. Stefan Institute, Jamova 39, P. O. Box 3000, 1001 Ljubljana, Slovenia}
\affiliation{Department of Physics, University of Ljubljana, Jadranska 19, 1000 Ljubljana, Slovenia}

\author{Jernej F. Kamenik}
\email[Electronic address:]{jernej.kamenik@ijs.si}
\affiliation{J. Stefan Institute, Jamova 39, P. O. Box 3000, 1001 Ljubljana, Slovenia}
\affiliation{INFN, Laboratori Nazionali di Frascati, I-00044 Frascati, Italy}

\date{\today}

\begin{abstract}
We consider chiral loop  corrections to the  scalar form factor in $B_q \to D_q l \nu_l$ decays. First we consider chiral corrections to the $1/m_Q$ suppressed operators and then we propose the procedure for the extraction of the relevant form factor using lattice QCD results. In the case of $B_s \to D_s l \nu_l$ decay we find that effects of kinematics and chiral corrections tend to cancel for the scalar form factor contributions. In particular the $1/m_Q$ suppression of chiral corrections is compensated by the  $SU(3)$ flavor symmetry breaking corrections which can be as large as $30\%$.
The calculated  corrections are  relevant for  the precise determination of possible new physics effects in $B_q \to D_q l \nu_l$ decays. 
\end{abstract}


\maketitle

\section{Introduction}

The $b\to c$ transitions have been traditionally important for the determination of the $V_{cb}$ CKM matrix element. For the exclusive determination at the B factories, the $B\to D^* e\nu$ decay channel is the most suitable. In addition to being the dominating B meson semileptonic decay mode, heavy quark symmetry (Luke's theorem) protects the normalization of the only relevant form factor at zero recoil point from he leading $1/m_Q$ power corrections~\cite{Luke:1990eg}. Furthermore, its shape can be extracted from the experimentally measured decay spectrum. However, with the modulus of $V_{cb}$ determined to a percent level, the focus is shifting toward constraining possible new physics contributions also in charged current transitions -- traditionally a domain of neutral current studies. $B\to D \tau\nu$ is one such promising process, where the large tau mass makes it sensitive to helicity suppressed contributions as such coming from the exchange of charged scalars~\cite{Tanaka:1994ay,Kiers:1997zt,Nierste:2008qe,Kamenik:2008tj}. This calls for high precision theoretical estimates for both the dominating vector ($G(w)$) as well as the relative scalar form factor contribution ($\Delta(w)$) describing the non-perturbative QCD dynamics of this transition. First such estimates on the lattice have already been performed~\cite{Hashimoto:1999yp,Okamoto:2004xg,deDivitiis:2007uk,deDivitiis:2007ui}. 

At the same time, high luminosity hadron machines (Tevatron and LHC) will take the central stage in the coming years also in the area of precision heavy flavor studies. Their higher energies allow to study processes, unavailable at the B factories. $B_s \to D_s$ transitions are examples of these. Although to our knowledge no non-perturbative lattice estimates exist at the moment for the required form factors, they are actually better suited for lattice studies than $B\to D$ transitions, as large chiral extrapolations to the physical light valence quark masses and the associated systematical errors can be largely avoided.

In this note we estimate the leading chiral symmetry breaking corrections, governing the differences between the vector and scalar form factors in $B_s\to D_s$ and $B\to D$ transitions. These corrections can on the one hand be used to guide lattice QCD studies in their chiral extrapolations. On the other hand we use them to estimate qualitatively the relative scalar form factor values in $B_s\to D_s$ transitions.

The leading order (LO) $SU(3)$ chiral corrections at leading and next to leading order (NLO) in $1/m_Q$ to the $B_q\to D_q^{(*)}$ semileptonic form factors have previously been computed in~\cite{Boyd:1995pq}. However phenomenological discussion at the time was limited by the lack of experimental and lattice information available on the sub-leading form factor contributions. We are now able to provide the first more reliable estimates of these contributions and also study their impact on $B_s\to D_s\tau\nu$ phenomenology.

\section{Framework}

\subsection{$1/m_Q$ power expansion of heavy quark currents}

The matrix element of the vector $b\to c$ quark current between a B and a D meson of velocity $v$ and $v'$ respectively, can be parametrized in terms of two form factors
\begin{eqnarray}
&& \bra{D(v')}\bar c \gamma_{\mu} b \ket{B(v)} = \sqrt{m_B m_D} \left[ h_+(w) (v+v')_{\mu}\right.  \nonumber\\ && \hspace{4.8cm} \left. + h_-(w) (v-v')_{\mu} \right]\,,\nonumber\\
\end{eqnarray}
where $w=v\cdot v'$. In heavy quark expansion they take the form~\cite{Falk:1992wt}
\begin{subequations}
\begin{eqnarray}
\label{eq:h+expansion}
h_+(w) &=& \xi(w) + \mathcal O (\epsilon_Q)\,,\\
\label{eq:h-expansion}
h_-(w) &=& (\epsilon_c-\epsilon_b) \left[-\bar \Lambda \xi(w) +2 \xi_3(w) \right]+ \mathcal O (\epsilon_Q^2)\,,\nonumber\\
&&
\end{eqnarray}
\end{subequations}
where $\epsilon_Q=1/2m_Q$ and $\bar\Lambda = m_{H_Q}-m_Q$. The $\xi(w)$ and $\xi_3(w)$ form factors are defined through the matching of the LO and NLO quark currents in HQET to HM$\chi$PT~\cite{Isgur:1989ed,Isgur:1990jf}. For the leading ($\xi(w)$) form factor (the Isgur-Wise function) one gets
\begin{eqnarray}
 \bar h' \Gamma h  &\to & -\xi(w) \mathrm{Tr} [ \bar H' \Gamma H ]\,,
\end{eqnarray}
where $h(x) = \exp{(-i m_Q v\cdot x)} P_+ Q(x)$ is the HQET heavy quark field, $P_+=(\gamma\cdot v + 1)/2$ is the HQET velocity-spin projection operator, while $H(x)=P_+[P^{*}(x)\cdot\gamma - P(x) \gamma_5]$ is the HM$\chi$PT $1/2^-$ field of vector and pseudoscalar mesons. All the primed (') quantities refer to the primed velocity $v'$ and flavor (mass $m_{Q'}$). The leading order contribution to the $h_-(w)$ form factor originates from the matching of QCD to NLO HQET currents. At tree level, the Lorentz structures of $\Gamma$ are the same on both sides and we are interested in the case $\Gamma=\Gamma_{V-A}^{\mu}\equiv\gamma_{\mu}(1-\gamma_5)$~\cite{Falk:1992wt}
\begin{subequations}
\begin{eqnarray}
 \bra{H'}\bar h' \Gamma \gamma^{\alpha} i \mathcal D_{\alpha} h\ket{H} & = & \mathrm{Tr} [  \bar H' \Gamma P_- \gamma^{\alpha} H \xi_{\alpha}(v,v')]\,, \nonumber\\
&&\\
 \bra{H'}(- i {\mathcal D_{\alpha}}) \bar h' \gamma^{\alpha} \Gamma h \ket{H}  & = &  \mathrm{Tr} [ \bar \xi_{\alpha}(v',v) \bar H'\gamma^{\alpha} P'_-\Gamma H ]\,,\nonumber\\
&&
\end{eqnarray}
\end{subequations}
where 
\begin{equation}
\xi_{\alpha}(v,v')=\xi_+(w)(v+v')_{\alpha} + \xi_{-}(v-v')_{\alpha}-\xi_3(w)\gamma_{\alpha}\,,
\end{equation} 
and $\bar \xi_{\alpha} = \gamma_0\xi_\alpha^{\dagger}\gamma_0$. The following identities can be proven: $\xi_-(w)=\bar \Lambda \xi(w)/2$, $\xi_+(w)(w+1)-(w-1)\xi_-(w)+\xi_{3}(w)=0$, reducing the number of independent form factors to the two ($\xi(w)$, $\xi_3(w)$) in eq.~(\ref{eq:h-expansion}).
Additional $1/m_Q$ form factors originate from the insertions of NLO HQET interaction Lagrangian. They contribute however only to $h_+(w)$ at $\mathcal O(\epsilon_Q)$ and as such do not interfere in the LO determination of $h_-(w)$. 
%$ \mathcal L_1 = \bar h (i D)^2 h + Z \bar h (-i/2) \sigma_{\alpha\beta} G^{\alpha\beta} h$ into the matrix elements of LO HQET %currents:
%\begin{subequations}
%\begin{eqnarray}
%i \int \mathrm{d}x T\{ (\bar h' \Gamma h)(0), \mathcal L_1(x) \} &\to & - A_1(w) \mathrm{Tr} \left[ \bar H' \Gamma H \right] \nonumber\\
%&&\hspace{-2.9cm}- Z \mathrm{Tr}\left[ A_{\alpha\beta} (v,v') \bar H' \Gamma P_+ (-i/2) \sigma^{\alpha\beta} H \right]\,,\nonumber\\
%&&\\
%i \int \mathrm{d}x T\{ (\bar h' \Gamma h)(0), \mathcal L'_1(x) \} &\to & - A_1(w) \mathrm{Tr} \left[ \bar H' \Gamma H \right]\nonumber\\ 
%&&\hspace{-2.9cm}- Z' \mathrm{Tr}\left[\bar A_{\alpha\beta} (v',v) \bar H'  (-i/2) \sigma^{\alpha\beta} P'_+ \Gamma  H \right]\,,\nonumber\\
%\end{eqnarray}
%\end{subequations}
%where $\sigma_{\alpha\beta}=i[\gamma_{\alpha},\gamma_{\beta}]/2$, $Z\equiv Z(m_Q/\mu)$ is the renormalization factor of the chromo-magnetic moment operator and
%\begin{equation}
%A_{\alpha\beta} (v,v') = A_2(w) (v'_{\alpha} \gamma_{\beta}-v'_{\beta} \gamma_{\alpha}) + A_3(w) i \sigma_{\alpha\beta}.
%\end{equation}

\subsection{Chiral expansion in HM$\chi$PT}

\begin{figure}[!t]
\begin{center}
\epsfysize2.3cm\epsffile{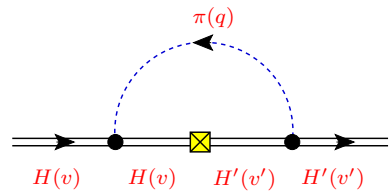}
\end{center}
\caption{\label{fig:sunrise}{Weak vertex correction ``sunrise'' diagram. Crossed box represents effective weak current operator vertex, while filled black circles represent effective strong vertices.}}
\end{figure}

The framework of the chiral heavy meson Lagrangians and procedures involved in the computation of loop corrections in the chiral expansion is by now standard and has been explained previously (c.f.~\cite{Eeg:2007ha}).
Novel however is the consideration of the  HM$\chi$PT Feynman rules for the sub-leading current operators from the previous section~\footnote{Our results agree with the convention of ref.~\cite{Boyd:1995pq} with the replacement $h_{5,6}\to h_+\mp h_-$}. Since all of them involve power corrections to both initial and final states,  their linear combinations entering expressions~(\ref{eq:h+expansion},\ref{eq:h-expansion}) in general mix under chiral corrections. The diagrams needed to be computed are the self-energy contributions for both the initial and final states as well as the ``sunrise'' diagrams with pseudo-Goldstone exchanges between the initial and final states connected to the weak current operators (fig.~\ref{fig:sunrise}). 
\psfrag{pi}[bl]{\footnotesize
$\Red{\pi(q)}$} \psfrag{Ha}[cc]{\footnotesize $\Red{H(v)}$}
\psfrag{Hb}[cc]{\footnotesize $\Red{H'(v')}$}
\psfrag{Hc}[cc]{\footnotesize $~\Red{H(v)}$}
\psfrag{Hd}[cc]{\footnotesize $~~\Red{H'(v')}$}
At LO in the chiral expansion no additional pseudo-Goldstone fields can be emitted from the weak current operator vertices. In the computation of the power suppressed chiral corrections to $h_+(w)$, one would need to consider contributions from the NLO HM$\chi$PT interaction Lagrangian~\cite{Boyd:1995pq}. These are however further suppressed when considering LO chiral corrections to the sub-leading form factor $h_-(w)$. 

In our phenomenological discussion we consider two cases: if one is to use the one-loop formulae to extrapolate lattice QCD results on the scalar form factor to the chiral limit, one has to use the $SU(2)$ theory. Namely, it has been shown previously that LO chiral symmetry breaking contributions in an $SU(3)$ limit cannot be reliably disentangled from the contributions of the lowest lying $1/2^+$ excited heavy meson states~\cite{Eeg:2007ha, Fajfer:2006hi, Becirevic:2006me}. On the other hand with a number of additional assumptions one can provide a qualitative estimate of the $B_s\to D_s$ sub-leading form factors even before lattice QCD results on these become available using $SU(3)$ theory predictions.

\section{Phenomenology}

The $B_q\to D_q \ell\nu$ partial rate in the SM can be written in
terms of $w=v_B\cdot v_D$ as 
\begin{eqnarray}
\label{eq:width}
      \frac{d\Gamma(B\to D\ell\overline{\nu})}{dw}
        &=& \frac{G_F^2|V_{cb}|^2 m_B^5
        }{192\pi^3}\rho_V(w)\\
         &\times&\left[1 -
      \frac{m_{\ell}^2}{m_B^2}\,
     \rho_S(w) \right] ,\nonumber
\end{eqnarray}
where $t(w) = m_B^2+ m_D^2 - 2w m_D m_B$ and we have decomposed
the rate into the vector and scalar Dalitz density contributions
\begin{eqnarray}
     \rho_V(w)  &=&
4\,\left(1+\frac{m_D}{m_B}\right)^2\left(\frac{m_D}{m_B}\right)^3\left(w^2-1\right)^{\frac{3}{2}}\nonumber\\
&&\times         \left(1-\frac{m_\ell^2}{t(w)}\right)^2
           \left(1+\frac{m^2_\ell}{2 t(w)}\right)\,G(w)^2,\,\quad\quad\\
    \rho_S(w) & =& \frac{3}{2}
    \frac{m_B^2}{t(w)}
         \,
           \left(1+\frac{m^2_\ell}{2 t(w)}\right)^{-1}
         \frac{1+w}{1-w}\,\Delta(w)^2\,.
\end{eqnarray}
Possible new charged current scalar interactions only contribute to $\rho_S$, introducing a shift $\rho_S(w) \to [1+ t(w) \delta_{NP}]\rho_S(w)$, where $\delta_{NP}$ parametrizes the new physics contribution~\cite{Kiers:1997zt,Nierste:2008qe,Kamenik:2008tj}. 
The $G(w)$ and $\Delta(w)$ form factors can be written in terms of $h_+(w)$ and $h_-(w)$ as
\begin{eqnarray}
	G(w) &=& h_+(w) - \frac{m_B-m_D}{m_D+m_B} h_-(w)\,\\
	\label{eq:delta}
\Delta(w) &=& \left[ \frac{m_B-m_D}{m_B+m_D} - \frac{w-1}{w+1} \frac{h_-(w)}{h_+(w)} \right]\nonumber\\
&&\times\left[ 1- \frac{m_B-m_D}{m_B+m_D} \frac{h_-(w)}{h_+(w)} \right]^{-1}\,.
\end{eqnarray}
The benefit of such form factor decomposition is also that $G(w)$ completely dominates $B\to D \ell\nu$ decay rates with $\ell= e,\mu$ and its shape can actually be extracted from the corresponding decay spectra. On the other hand, $\Delta(w)$ is only relevant for $B\to D \tau\nu$ and in addition depends on the non-perturbative QCD dynamics only through the ratio of form factors $h_-/h_+$, making it highly suitable for lattice QCD studies.

$G(w)$ receives chiral corrections already at LO in $1/m_Q$ (the well known chiral corrections to the Isgur-Wise function $\xi(w)$~\cite{Jenkins:1992qv, Cho:1992cf}) making the sub-leading contributions perhaps less important. Only at zero recoil ($w=1$), Luke's theorem protects $h_+(w)$ from $1/m_Q$ corrections ($\xi(1)=1$ by construction and thus invariant under chiral corrections), and here the leading order chiral corrections to $G(1)$ actually come from $1/m_Q$ suppressed $h_-(w)$ contributions. The case is similar for $\Delta(w)$, whose leading order value is strictly determined by kinematics throughout the kinematic region. Here the chiral corrections to sub-leading terms (proportional to $h_-/h_+$) are dominating the chiral extrapolation of lattice studies as well. Both cases underline the importance of controlling the chiral corrections of the $h_-(w)$ form factor. We turn to this issue first.

\subsection{Lattice QCD extrapolation formulae}

Below we give the leading logarithmic chiral corrections relevant for lattice QCD extraction of the $h_+(w)$ and $h_-(w)$ form factors in the $N_f=2$ theory ($r(x) = \log(x+\sqrt{x^2-1})/\sqrt{x^2-1}$)
\begin{subequations}
\begin{eqnarray}
\label{eq:hpchi}
	h_+ &=& h_+^{\mathrm{Tree}} \left[ 1 + 3 g^2 \frac{r(w)-1}{(4\pi f_{\pi})^2}  m_{\pi}^2\log \frac{m_{\pi}^2}{\mu^2} + m_{\pi}^2 c_{+}(\mu,w) \right]\,, \nonumber\\
&&\\
\label{eq:hmchi}
	h_- &=&  h_-^{\mathrm{Tree}} \left\{ 1 - 3 g^2 \frac{2 + Y_+ [r(w)+1]}{(4\pi f_{\pi})^2}  m_{\pi}^2\log \frac{m_{\pi}^2}{\mu^2} \right.\nonumber\\
&&\hspace{4.5cm}+ m_{\pi}^2 c_{-}(\mu,w) \Big\}\,,
\end{eqnarray}
\end{subequations}
where $Y_+=(\epsilon_c-\epsilon_b) \bar \Lambda h_+^{\mathrm{Tree}}/h_-^{\mathrm{Tree}}$ and the $w$ dependence is implicit. Here and in the rest of the text $c_i(\mu,w)$ denote the sums of local analytic counter-terms, which cancel the $\mu$ dependence of the chiral log pieces. As stressed by the notation, they will in general have a non-trivial $w$ dependence, originating both from the finite analytic residuals of chiral loops as well as from local NLO chiral current operators, needed to cancel the UV divergences of chiral loops. We have used eqs.~(\ref{eq:h+expansion},\ref{eq:h-expansion}) to rewrite $\xi_{(3)}(w)$ in terms of $h_{\pm}^{\mathrm{Tree}}(w)$, which is valid up to $1/m_Q$ corrections. Since $r(1)=1$, chiral corrections to $h_+(1)$ at LO (and also at NLO~\cite{Boyd:1995pq}) in $1/m_Q$ vanish as they should, and the dominating contributions come at $1/m_Q^2$~\cite{Randall:1993qg}. The value of $h_-(1)$ however is not protected by heavy quark symmetry. Note also that the one-loop chiral corrections to $h_-(w)$ are proportional to a different combination of $\xi_3(w)$ and $\xi(w)$ than the tree level value resulting in an admixture of $h_+(w)$ in the $\mu$ of $m_{\pi}$ running of $h_-(w)$. This can be most easily understood at zero recoil, where the value of $\xi(1)$ admixture in $h_-(1)$ is protected from receiving chiral corrections by heavy quark symmetry. Therefore they should be proportional exclusively to $\xi_3(1)$, which is the case. Actually the eigenvectors of the two coupled equations~(\ref{eq:hpchi},\ref{eq:hmchi}) in the whole kinematic region are precisely $\xi(w)$ and $\xi_3(w)$. For completeness, we give their explicit chiral LO corrections formulae in the appendix. Incidentally, when evaluating the $\Delta(w)$ form factor, only the ratio $h_-/h_+$ is needed, whose chiral corrections take on a simpler form, as can be easily inferred by dividing eqs.~(\ref{eq:hpchi},\ref{eq:hmchi}) and then expanding up to linear terms in $m^2_{\pi}\log m^2_{\pi}$
\begin{eqnarray}
	\frac{h_-}{h_+} &=&  \left(\frac{h_-}{h_+}\right)^{\mathrm{Tree}} \left[ 1 - 3 g^2 \frac{r(w)+1}{(4\pi f_{\pi})^2}  m_{\pi}^2\log \frac{m_{\pi}^2}{\mu^2}  \right]\nonumber\\
&&\hspace{-1.3cm}- \bar\Lambda (\epsilon_c-\epsilon_b) 3 g^2 \frac{r(w)+1}{(4\pi f_{\pi})^2}  m_{\pi}^2\log \frac{m_{\pi}^2}{\mu^2} + m_{\pi}^2 c_{{\pm}}(\mu,w)\,.\nonumber\\
\end{eqnarray}
Again one can understand the form of chiral corrections by rewriting the ratio $h_-/h_+ \to (\epsilon_c-\epsilon_b)(-\bar \Lambda + 2\xi_3(w)/\xi(w))$. Since $\bar\Lambda$ is defined independently of spectator effects, it should not receive chiral corrections. Therefore they should again be proportional to $\xi_3(w)/\xi(w)$ {\it throughout the kinematic region}, as can be readily checked by using eqs.~(\ref{eq:h+expansion},\ref{eq:h-expansion}). 

The same sub-leading IW functions enter also other $b\to c$ hadronic matrix elements. In particular, the $\bar\Lambda (\epsilon_c-\epsilon_b)$ contribution introducing explicit heavy quark mass dependence or equivalently, the mixing of leading and subleading form factor contributions can be avoided by rewriting the chiral extrapolation formula (\ref{eq:hmchi}) in terms of the tree level matrix element $\bra{D^*(\epsilon',v')}\Gamma\ket{B^*(\epsilon,v)}$ part proportional to $-\epsilon\cdot\epsilon'(v-v')$ -- we denote it with $h^*_-$~\footnote{In the notation of ref~\cite{Boyd:1995pq} $h^*_-=(f_4-f_5)/2$, while in convention of ref.~\cite{Falk:1992wt} $h^*_=h_2$}. In heavy quark expansion it can be written as $h^*_- (w)= - \bar\Lambda (\epsilon_c-\epsilon_b) \xi(w)  + \mathcal O(\epsilon_Q^2)$~\cite{Falk:1992wt}. Immediately we see, that its leading chiral corrections should coincide with those of $h_+$ as can also be checked by computing the two sunrise diagrams contributing to this vertex correction. The chiral extrapolation equation for $h_-(w)$ (\ref{eq:hmchi}) on the other hand remains of the same form with the replacement $- Y_+ \to Y^*=(h^*_-/h_-)^{\mathrm{Tree}}$. 
The goal of lattice simulations is then to determine $h_{\pm}^{\mathrm{Tree}}(w)$ or alternatively ${h^{(*)\mathrm{Tree}}_-}(w)$, and $c_{i}(w)$ by fitting the above formulae to the numerical values of $h_{\pm}(\{m_{\pi}\})(w)$, $h^*_-(\{m_{\pi}\})(w)$ at several values of $m_{\pi}$ for each considered $w$. Using eqs.~(9) and (10) (and (7) and (8)) one can thus obtain all physically relevant hadronic variables needed in the search for possible charged current scalar interactions in $B_q\to D_q\tau\nu$.

\subsection{$B_s\to D_s$ estimate for $\Delta(w)$}

\begin{figure}[t]
\begin{center}
\end{center}
\scalebox{0.7}{\includegraphics{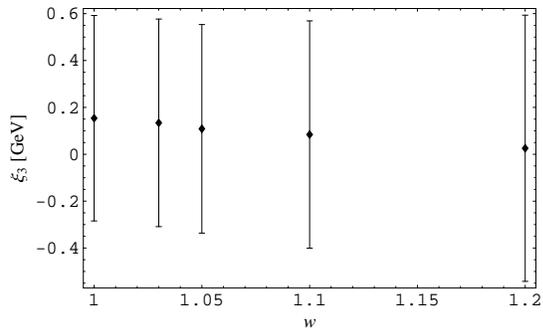}}
\caption{\label{fig:xi_3}Estimate of $\xi_3(w)$ from lattice data on $h_{\pm}(w)$~\cite{deDivitiis:2007uk,deDivitiis:2007ui}.}
\end{figure}

\begin{figure}[t]
\begin{center}
\end{center}
\scalebox{0.7}{\includegraphics{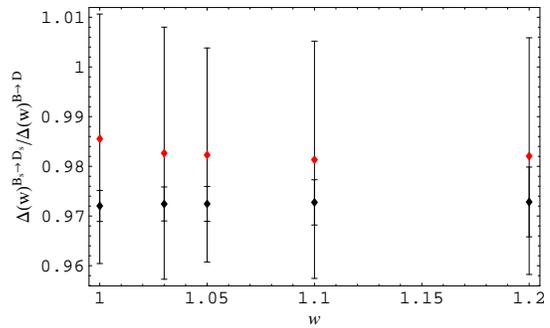}}
\caption{\label{fig:deltasd}Estimate of $\Delta^{B_s\to D_s}(w)/\Delta^{B\to D}(w)$ using $N_f=3$ logarithmic chiral corrections and information from lattice data on $G(w)$ and $\Delta(w)$. The leading order kinematic effect is shown in black, while chirally corrected result is shown in red. Only uncertainties coming from lattice inputs~\cite{deDivitiis:2007uk,deDivitiis:2007ui} are taken into account in the error bars displayed. }
\end{figure}

We can revert the game and use the already available information on the $h_-(w)$ form factor in $B\to D$, to estimate the size of the sub-leading form factor $\xi_3(w)$. Then we can use this information together with a number of assumptions to estimate $\Delta(w)$ in $B_s\to D_s$. Namely, using eqs.~(\ref{eq:h+expansion},\ref{eq:h-expansion}) with $\bar\Lambda \simeq 0.5$~GeV~\cite{Neubert:1992fk} and the $\overline{\mathrm{MS}}$ charm and $b$ quark masses $m_c(m_c)=1.2$~GeV, $m_b(m_b)=4.2$~GeV~\cite{Yao:2006px} we can extract from available lattice data on $h_+(w)$ and $h_-(w)$~\cite{deDivitiis:2007uk,deDivitiis:2007ui} the size of $\xi(w)$ at leading order in $1/m_Q$ and consequently $\xi_3(w)$ for the calculated lattice points as shown in figure~\ref{fig:xi_3}.
Then we use the computed chiral corrections to $\xi(w)$ and $\xi_{3}(w)$ in the $N_f=3$ limit, neglect the analytic terms at a fixed scale of $\mu=1$~GeV and also neglect contributions from excited heavy meson states to obtain an estimate for $\Delta(w)$ in $B_s\to D_s$. The dominating contribution naturally comes from the LO kinematic factor since $m_{D_s}/m_{B_s}\neq m_{D}/m_B$. This contribution amounts to roughly 3\% supression from unity in the ratio $\Delta^{B_s\to D_s}(w)/\Delta^{B\to D}(w)$. Additional chiral contributions come from the 
%difference in $\bar\Delta_{s,q}$ and from the 
 chiral corrections to the HQET form factors entering $\Delta(w)$ and can be put into the approximate form 
\begin{eqnarray}
\frac{h_-^s h_+^d}{h_-^d h_+^s} &=&  1 + 2 g^2 \frac{(1-Y_{\xi})(1+r(w))}{(4\pi f_{\pi})^2} \nonumber\\
&&\hspace{-1.5cm}\times \left( \frac{3}{2} m_{\pi}^2\log \frac{m_{\pi}^2}{\mu^2} -  m_{K}^2\log \frac{m_{K}^2}{\mu^2} - \frac{1}{2} m_{\eta}^2\log \frac{m_{\eta}^2}{\mu^2} \right) + \mathrm{c.t.}\,,\nonumber\\
\label{eq:fdfs}
\end{eqnarray}
where $1/Y_{\xi}=1-2\xi_3/\bar\Lambda\xi$ and c.t. denotes additional counter-term contributions. 
Both kinematic and chiral log effects are shown in fig.~\ref{fig:deltasd}~\footnote{Note that in our numerical analysis we do not use the approximate expanded formula~(\ref{eq:fdfs}) but insert chiral corrections formulae~(\ref{eq:hpchi},\ref{eq:hmchi}) into the ratio of $\Delta(w)$s directly. We checked however that the ratio values obtained in this way do not differ significantly from the ones obtained with the approximation~(\ref{eq:fdfs}).}. Only uncertainties coming from lattice inputs are taken into account in the error bars displayed. 
We see that the effects are comfortably small, point at suppression smaller than 5\%, and are consistent with unity ratio within the current large errors throughtout the available $w$ region. The chiral corrections are supressed by $(2\xi_3-\bar \Lambda \xi)(\epsilon_c-\epsilon_b)\sim 0.1$. However, this is compensated by the large $SU(3)$ breaking corrections which are of the order $30\%$. A more reliable estimate is of course expected from the lattice. 
%In passing we also note that the $w$ dependence of the ratio in eq.~(\ref{eq:fdfs}) and as seen on fig.~(\ref{fig:deltasd}) signifies that one should expect deviations from the approximately constant behavior of $\Delta(w)$ observed in present lattice results on $B\to D$ transitions~\cite{deDivitiis:2007uk,deDivitiis:2007ui}, since it is not invariant under chiral corrections. In addition a na\"ive constant extrapolation of the central value of $\Delta$ to the  (unphysical) $t(w)=0$ point would violate the kinematical constraint (formally satissfied by construction) $\Delta(t(w)=0)=(m_B-m_D)/(m_B+m_D)$.

%As a final application, we can use the available information on the $G$ and $\Delta$ form factors from $B\to D$ transitions together with the leading chiral $SU(3)$ theory formulae derived above to estimate the ratios $Br(B_s\to D_s \ell\nu)/Br(B\to\D\ell\nu)$ and $Br(B_s\to D_s \tau\nu)/Br(B_s\to D_s \ell\nu)$, which are both free from uncertainties due to $|V_{cb}|$ and the second also from form factor normalizations.

\section{Conclusions}

Motivated by the possible sensitivity to the presence of new physics in
$B_q \to D_q \tau \nu$ decays~\cite{Tanaka:1994ay,Kiers:1997zt,Nierste:2008qe,Kamenik:2008tj} we have investigated the effects of chiral corrections to the relevant (scalar) form factor.
The explicit chiral behavior of the computed chiral corrections can be used to guide future lattice computations in approaching the physical regime for the light quark masses.

Based on results of our previous studies~\cite{Eeg:2007ha, Fajfer:2006hi, Becirevic:2006me} we calculate the leading chiral logarithmic behavior
of the $h_-(w)$ and $h_+(w)$ form factors, by including only pionic contributions.
We consider chiral corrections of the operators appearing at $1/m_Q$ in matching QCD to HQET and which contribute to the scalar form factor. We propose to determine the matrix elements for the $B \to D$ and $B^* \to D^*$ transitions with the same kinematics dependence ($v-v'$) on the lattice simultaneously. In this way one can extract the scalar form factor at tree level in the chiral expansion, as well as the relevant counter-term needed to make predictions in the physical regime. 
%Eventually, the reliability of the chiral extrapolation formulae for $h_-(w)$ or $\Delta$ could be improved by including NLO $1/m_Q$ power suppressed contributions as already done for the $F(w)$ form factor in $B\to D^*\ell\nu$ transitions~\cite{Hashimoto:2001nb}.

Finally we consider chiral corrections to the scalar form factor in the $B_s \to D_s$ transition and
conclude that $SU(3)$ flavor symmetry breaking effects are significant (up to $30 \%$) and can thus partly compensate the $1/m_Q$ suppression of chiral corrections. They are comparable in size with kinematical effects but tend to be of the opposite sign. This points towards longitudinal/scalar contributions very similar to $B \to D$ transitions.

 \begin{acknowledgments}
 This work is supported in part by the EU-RTN Programme,
Contract No. MRTN--CT-2006-035482, \lq\lq Flavianet'', by the Norwegian Research Council and by the Slovenian Research Agency.
 \end{acknowledgments}

\appendix

\section{$SU(N)$ chiral corrections formulae for $\xi(w)$ and $\xi_3(w)$}

Below we give the complete expressions of the chiral 1-loop corrected
form factors $\xi(w)$ and $\xi_3(w)$, which are eigenvectors of the chiral running and are needed for the LO lattice chiral extrapolation of $h_{\pm}(w)$. They are given in a $SU(N)$ theory where $t_i$ denotes the $SU(N)$ generator matrices and a summation of the repeated index $i$ is assumed. As in the text $r(x) = \log(x+\sqrt{x^2-1})/\sqrt{x^2-1}$.
\begin{subequations}
\begin{eqnarray}
	\xi(w) &=& \xi(w)^{\mathrm{Tree}} \left\{ 1 +  t_i t_i^{\dagger}  \frac{2 g^2[r(w)-1]}{(4\pi f_{\pi})^2}  m_{i}^2\log \frac{m_{i}^2}{\mu^2} + 
\mathrm{c.t.}\right\}, \nonumber\\
&&\\
	\xi_3(w) &=& \xi_3(w)^{\mathrm{Tree}} \left[ 1 -  t_i t_i^{\dagger}  \frac{4 g^2}{(4\pi f_{\pi})^2}  m_{i}^2\log \frac{m_{i}^2}{\mu^2} +
\mathrm{c.t.} \right]. \nonumber\\
&&
\end{eqnarray}
\end{subequations}
Here c.t. denotes the corresponding counter-terms, which cancel the $\mu$ dependence of the log terms. The counter-terms are analytic in $m_i$ and will generally themselves have a non-trivial $w$ dependence, so they should be fitted for each value of $w$ independently. This is also true for $\xi_3(w)$ despite the fact, that its LO non-analytic chiral corrections are $w$ independent. 
%As explained in the text, $\xi(w)$ can be extracted on the lattice at LO in $1/m_Q$ from the equality with $h_+(w)$ (but equally also from $h_{A_1}$ in $B\to D^*$ or $h_1$ in $B^*\to D^*$ transitions in notation of reference~\cite{Falk:1992wt}). On the other hand $\xi_3$ cannot be so easily disentagled from other contributions with the same kinematics. We find however, that the value of $\xi_3(1)$ is at LO in $1/m_Q$ proportional to the $h_3(1)-h_4(1)$ form factor combination in $B^*\to D^*$ transitions, again in notation of reference~\cite{Falk:1992wt}). 

\bibliography{article}

\end{document}